\documentclass[twocolumn,showpacs,preprintnumbers,amsmath,amssymb]{revtex4}

\usepackage{graphicx}
\usepackage{dcolumn}
\usepackage{bm}

\begin{document}

\preprint{}

\title{Cosmic-ray electron signatures of dark matter}

\author{Martin Pohl}
\email{mkp@iastate.edu}
\affiliation{Department of Physics and Astronomy, Iowa State University, Ames,
IA 50011
}%

\date{\today}

\begin{abstract}
There is evidence for an excess in cosmic-ray electrons at about 500~GeV energy,
that may be related to dark-matter annihilation. I have calculated 
the expected electron contributions from a pulsar and from Kaluza-Klein dark matter, based on
a realistic treatment of the electron propagation in the Galaxy.
Both pulsars and dark-matter clumps are quasi-pointlike and few, and 
therefore their electron contributions at Earth generally have spectra that deviate from the 
average spectrum one would calculate for a smooth source distribution. I find that
pulsars younger than about $10^5$~years naturally cause a narrow peak at a few hundred GeV
in the locally observed electron spectrum, similar to that observed. On the other hand,
for a density $n_c = 10\ {\rm kpc^{-3}}$ of dark-matter clumps
the sharp cut-off in the contribution from Kaluza-Klein particles is sometimes more pronounced,
but often smoothed out and indistinguishable from a pulsar source, and therefore 
the spectral shape of the electron excess is insufficient to discriminate
a dark-matter origin from more conventional astrophysical explanations.
The amplitude of variations in the spectral feature caused by dark matter predominantly
depends on the density of dark-matter clumps, which is not well known.
\end{abstract}

\maketitle

\section{\label{sec:level1}Introduction}
Only about 1 per cent of galactic cosmic rays are electrons, but their
properties are of particular interest because they are very radiative at high energies
and thus quickly loose their energy. The cosmic-ray electron spectrum is therefore softer than
that of cosmic-ray nucleons, and high-energy electrons are few. For a long time
only emulsion-chamber data of the electron flux above 100~GeV were available \cite{elec}, which were
well represented by a power law $N(E)\propto E^{-3.2}$, but the energy resolution and
statistical accuracy were limited. Recently data obtained with the ATIC balloon experiment
were published, which show an excess of galactic cosmic-ray electrons at energies
between 300 and 700~GeV \cite{chang}. At TeV energies the electron flux appears to drop off rapidly 
\cite{hess}.
This excess is indicative of a previously unknown individual source
of high-energy electrons, which could be a nearby supernova remnant \cite{koba}, a pulsar \cite{aha},
a microquasar \cite{heinz}, or an annihilation site of dark-matter particles of the Kaluza-Klein
type \cite{kk}.

The dark-matter interpretation is particularly appealing, because the PAMELA
collaboration has reported an increase in the cosmic-ray positron fraction above 20~GeV,
suggesting the
existence of a local source of both positrons and electrons \cite{pam}. Dark-matter annihilation
resulting in electron-positron pairs is possible in a number of models \cite{cholis,berg,cirelli,arkani},
and will produce a spectrum dominated by a delta functional
at the mass of the dark-matter particle \cite{hs}. 
Indeed, the ATIC team finds that a Kaluza-Klein particle with mass 620~GeV fits their electron data
just fine,
when the expected electron source spectrum is propagated in the Galaxy using the GALPROP code \cite{mos}.
Hall and Hooper suggest that high-precision measurements of the electron spectrum 
be conducted with atmospheric Cherenkov observatories such as VERITAS and HESS, that would
permit a discrimination between the dark-matter and pulsar hypotheses \cite{hh}.

If dark-matter annihilation is responsible for the excess in 600-GeV electrons, then a substantial
boost factor is required to match the observed electron flux which requires that the dark matter
be concentrated in dense clumps. The electrons would then be injected into the Galaxy only at
the location of those clumps, which introduces substantial variations in the electron flux throughout
the Galaxy and can significantly modify the observed electron spectrum, as was shown in
similar studies of electron propagation from supernova remnants \cite{cl,pe98,p03}. The
GALPROP code implicitely assumes a smooth source distribution on account of its using
a finite-difference algorithm on a coarse grid, and therefore it will not properly describe
those fluctuations.

Here we study the propagation of relativistic electrons from localized sources. 

\section{\label{sec:model}The propagation of relativistic electrons}
The effects of the spatial structure of the electron sources appear only at
higher particle energies, at which the radiative loss time is short. 
Therefore we may treat
the propagation of electrons at energies above 50 GeV with
a simplified transport equation,
\begin{equation}
\frac{\partial N}{\partial t}- \frac{\partial}{\partial E} (bE^2\, N)
-D\,E^a\, {\bf\nabla}^2 N = Q
\label{transport}
\end{equation}
with which we consider continous energy losses by synchrotron radiation and inverse 
Compton scattering, a diffusion coefficient $D\,E^a$ dependent on energy, and 
a source term $Q$. Throughout this paper the propagation parameters have the values
\begin{equation}
{1\over {bE}}= 2.6\cdot 10^{15}\ {\rm s} , 
\end{equation}
corresponding to synchrotron losses in a $11\ {\rm \mu G}$ galactic magnetic field and Compton
scattering of the CMB. The diffusion coefficient is chosen as required in 
cosmic-ray propagation without continuous reacceleration \cite{j01,sm98}
\begin{equation}
D\,E^a= \left(10^{28}\ {\rm cm^2\,s^{-1}}\right)\,\left({E\over {\rm GeV}}\right)^{0.6}
\end{equation}
Green's function for this problem is \cite{gs64}
\begin{equation}
G={{\Theta(t-t')\,\delta\left( t-t' +{{E-E'}\over {b\, E\, E'}} \right)}
\over {bE^2\, \left( 4\pi\,\lambda\right)^{3/2} }}\ 
\exp \left(-{{({\bf r}-{\bf r'})^2}\over {4\, \lambda}}
\right)
\end{equation}
where $\Theta$ is a stepfunction and
\begin{equation}
\lambda= {{D\,\left(E^{a-1} -E'^{a-1}\right)}\over {b\,(1-a)}}
\end{equation}
In the case of discrete sources the injection term $Q$ is a sum over all such
sources. For an individual source we can write
\begin{equation}\label{eq:source}
Q_i=q_0\,f(E)\, g(t)\,\delta(r)
\end{equation}
and obtain the current ($t=0$) contribution of that source to the electron density at distance $r$ as
\begin{eqnarray}
N_i & = & {{q_0}\over {bE^2\,(4\pi)^{3/2}}}\,\int_{-\infty}^0 dt'\ g(t')\,
\delta\left(-t' +{{E-E'}\over {b\, E\, E'}} \right)\nonumber \\
& & \times \ \int dE'\ {{f(E')}\over {\lambda^{3/2}}}\,
\exp \left(-{{r^2}\over {4\, \lambda}}\right)
\end{eqnarray}

\subsection{\label{ssec:model:pulsar}Electrons from pulsars}
The spectrum of electron escaping from a pulsar is not well known. A simple parametrization may be in
order \cite{hh} that describes the differential production rate of electrons by a pulsar
\begin{equation}
f_p(E)=E^{-1.5}\,\exp\left(-{E\over {E_c}}\right)
\label{eq:source:pul}
\end{equation}
\begin{equation}
g_p(t)=\Theta\left(t+\tau\right)
\end{equation}
where we use a stepfunction to account for the finite age, $\tau$, of the pulsar.
Obviously, the spectrum of the excess electrons depends chiefly on the scaling parameters
$\xi=bE_c\,\tau$ and $\rho=(r^2\,\left[1-a\right]\,b)/(4\, D\, E_c^{a-1})$, 
which compare the age, $\tau$, with the
energy-loss timescale,$1/(b\,E_c)$, and the distance, $r$, with the diffusion length within one
energy-loss time.
At time $t=0$ and distance $r$ we observe the differential density of electrons as
\begin{eqnarray}
N_p & = & {C\over {E^{1+1.5a}}}\,\left({\rho\over {r^2}}\right)^{3\over 2}\,
\int_1^{x_{\rm max}} dx\ {{x^{-1.5}}\over {\left[1-x^{a-1}\right]^{3/2}}}\nonumber \\
 & & \times\ \exp\left(-x\,{E\over {E_c}}-\left({E\over {E_c}}\right)^{1-a}\,
{\rho\over {1-x^{a-1}}}\right)
\label{eq:np}
\end{eqnarray}
where $C$ absorbs all constants and 
\begin{equation}
x_{\rm max}=\begin{cases} \infty & \text{if ${E\over {E_c}}\,\xi\ge 1$,}\\
 & \\
{1\over {1-{E\over {E_c}}\,\xi}} & \text{if ${E\over {E_c}}\,\xi< 1$.}
\end{cases}
\end{equation}

\begin{figure}
\includegraphics[width=0.48\textwidth]{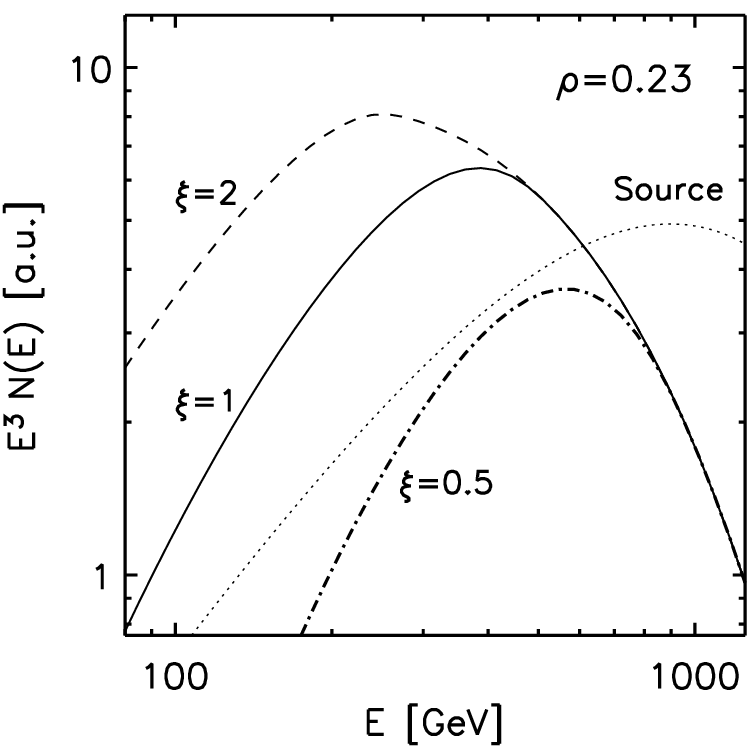}
\caption{\label{fig:1} Examples of electron spectra measured near Earth for different ages of
the pulsar (cf. Eq.~\ref{eq:np}). The thin dotted line denotes the pulsar source spectrum
as in Eq.~\ref{eq:source:pul}. The three thick lines give the particle spectra for three different
values ages parameter $\xi$, but the same distance parameter, $\rho$. }
\end{figure}
In Fig.~\ref{fig:1} we show possible electron spectra near Earth, that may result from a pulsar
that produces electrons with spectrum (\ref{eq:source:pul}), indicated by the thin
dotted line.
To be noted from the figure is the sharp peak in the electron spectrum that is 
entirely a propagation effect.
The distance to the pulsar is assumed as $r=700$~pc; for our propagation parameters
the distance parameter is then $\rho=0.23$, meaning electrons observed at $600$~GeV can well
reach Earth within one energy-loss time, whereas electrons beyond a few~TeV can not.

If $\xi=1$, the pulsar age is the same as the energy-loss time at 
600~GeV, $140,000$~years for our propagation parameters. Electrons at lower energy may not have
enough time to not reach Earth because their propagation range scale $\propto E^a$,
thus causing a sharp low-energy cut-off in the observed spectrum.
Given the local supernova rate we expect about one pulsar born within $1$~kpc and
$10^5$~years, so given the energy-loss time at 600~GeV one pulsar at $r=700$~pc is realistic. 
However, the pulsars need a few million years to propagate 1~kpc above or below the
galactic plane, and so electron injection by the pulsar must taper off after about $10^5$~years
to avoid a high flux of 100-GeV electrons.

\section{Electrons from dark matter}\label{sec:dm}
Having established in the preceding section that for realistic parameters a pulsar can produce
a narrow peak in the local electron flux, I will now discuss electron spectra that may arise from
dark-matter annihilation. As substantial boosting factors of a few hundred are needed \cite{cho08},
the dark matter is most likely organized in a number of individual high-density clumps. The size
of the clumps is not relevant for us, as long as it is much smaller than the propagation range
of 600-GeV electrons, a few hundred parsec. Likewise, the large-scale distribution of the clumps
in the Galaxy does not matter, unless the clump density varies on scales similar to the electron
propagation range. I will therefore assume the clumps to be randomly distributed in space with constant
density $n_c$. Dark-matter annihilation should proceed at a constant rate and can, in the case of 
Kaluza-Klein particles, produce electrons with a source spectrum that is dominated by 
a delta-functional at the particle mass. For each clump, the differential source rate of electrons
can then be described by the functions (cf. equation(\ref{eq:source}))
\begin{equation}
f_{dm}=\delta(E-E_c)\, ,\qquad g_{dm}(t)=1
\label{eq:source:dm}
\end{equation}
The electron spectrum observed at distance $r$ from the clump is then
\begin{equation}
N_{dm}={{C\,\Theta(E_c-E)}\over {E^2\,\lambda^{3/2}}}\,
\exp\left(-{{r^2}\over {4\, \lambda}}\right)
\label{eq:spec:dm}
\end{equation}
where $C$ absorbs the constants, $\Theta$ is a stepfunction, and 
\begin{equation}
\lambda={{D\,E_c^{1-a}}\over {b\,(1-a)}}\,\left[\left({{E_c}\over E}\right)^{1-a} -1\right]
\end{equation}
The total electron spectrum is then obtained by summing the contributions from all clumps.
If the clump density $n_c$ is high, the dark-matter distribution is effectively homogenous.
Then the total electron spectrum is governed by a cooling tail.
\begin{equation}
N_{dm,tot}=n_c\,\int_0^\infty dr\ 4\pi\,r^2\,N_{dm}=C'\,{{\Theta(E_c-E)}\over {E^2}}
\label{eq:galprop}
\end{equation}
This is the case implicitely (and tacitly) assumed when using the standard GALPROP code, and
it is presented in many publications \cite[e.g.][]{chang}.

\begin{figure}
\includegraphics[width=0.48\textwidth]{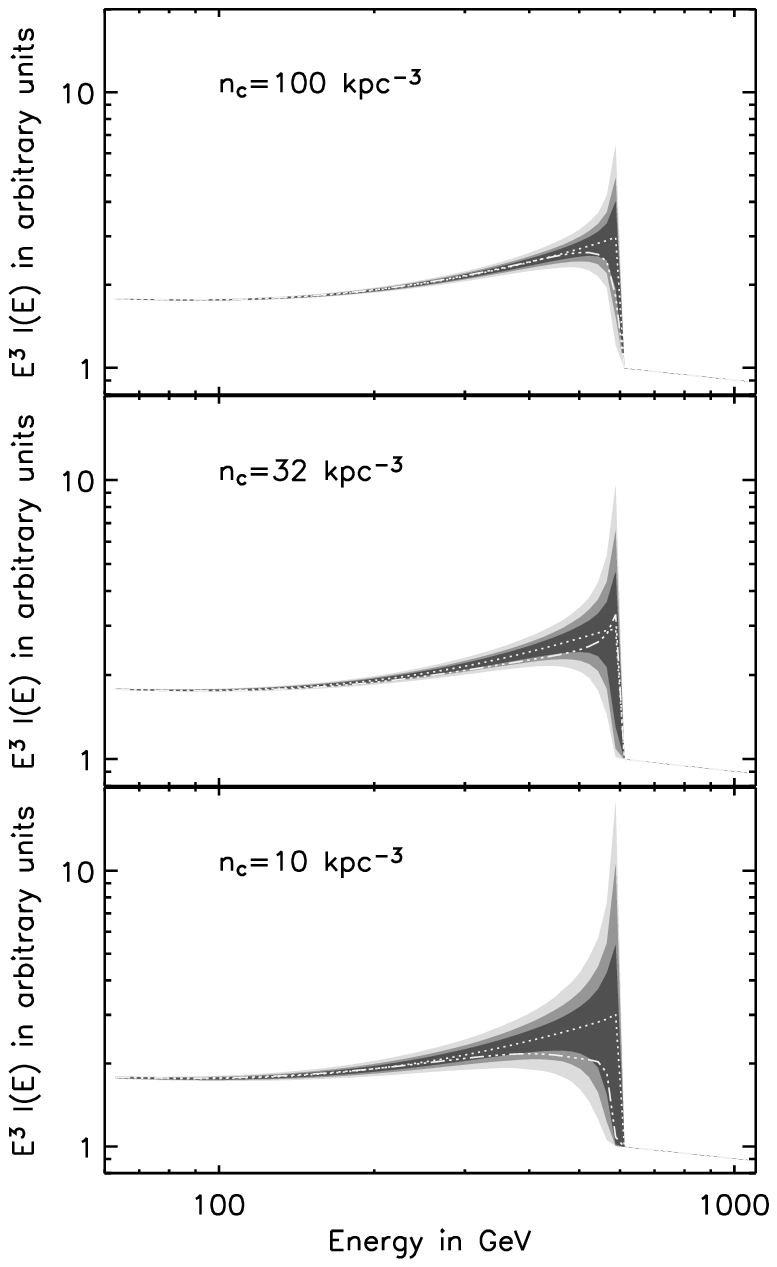}
\caption{\label{fig:2} The range of electron spectra measured near Earth for different
clump densities of dark matter, added onto the galactic electron background with spectrum
$\propto E^{-3.2}$. The dotted line indicates the mean spectrum (cf. Eq.~\ref{eq:galprop}).
The dash-dotted line denotes a randomly selected spectrum as an example of what may be observed.
The shaded areas indicates the range in which we find the electron flux in 68\% (dark gray),
90\% (medium gray), and 99\% (light gray) of all cases.}
\end{figure}
The question arises at what density the spectra start deviating from that for
the homogeneous case (Eq.\ref{eq:galprop}) and what the observed spectral shape might be.

I have randomly placed in the Galaxy dark-matter clumps with constant density $n_c$
and summed their electron contribution according to equation (\ref{eq:spec:dm}), using 
$E_c=600$~GeV. The resulting 
electron spectrum from dark matter is added to the generic galactic electron flux, for which
a spectrum $\propto E^{-3.2}$ is assumed. On average, the dark-matter component at 600~GeV has 
twice the flux of the galactic electron background.

Figure \ref{fig:2} shows the resulting total electron spectra for three different clump densities.
In all three panels the dotted line indicates the mean spectrum according
to Eq.~(\ref{eq:galprop}), and the dash-dotted line denotes a randomly selected spectrum out of the
5000 that were calculated. To be noted from the figure is that the dark-matter hump often doesn't 
look different from the hump a pulsar would produce. It can be fairly roundish and lack the sharp 
cut-off at $E_c$. The reason is that electrons at an energy very close to $E_c$ must be very 
young, because they haven't lost a significant fraction of their energy, and can therefore only 
come from a very close dark-matter clump. Even for $n_c=100\ {\rm kpc^{-3}}$ we expect only one 
clump within 140~parsec, and thus electrons at $E\simeq E_c$ may not reach us. On the other hand,
a very close dark-matter clump would produce a dominant spike at $E\simeq E_c$.

The shaded areas indicate the range of flux for three different probabilities. In 68\%
of all cases the electron flux is within the dark gray region, the medium gray area corresponds to 
90\% and the light gray to 99\% probability.

\section{Summary and discussion}
The ATIC collaboration has measured an excess in cosmic-ray electrons at about 500~GeV energy
\cite{chang}, which may be related to dark-matter annihilation. In this paper I have calculated 
the expected electron contributions from a pulsar and Kaluza-Klein dark matter. My emphasis is on
a realistic treatment of the electron propagation in the Galaxy, for which I use analytical 
solutions to the electron transport equation. The commonly employed GALPROP code implicitely 
assumes a smooth distribution of the electron sources, because it uses a finite-difference
algorithm on a grid.

The findings can be summarized as follows:
\begin{itemize}
\item Pulsars younger than about $10^5$~years naturally cause a narrow peak at a few hundred GeV
in the locally observed electron spectrum. A single pulsar could therefore explain both the
electron excess measured with ATIC and a similar excess in positrons, evidence for which at 50 to
100~GeV was obtained by the PAMELA experiment \cite{pam}. The pulsar hypothesis does require
that pulsars with ages $10^5$ to $10^6$ years leak significantly fewer electron/positron pairs,
otherwise they would provide a very strong contribution in the 50 to 300~GeV band that is not 
observed. 
\item Dark-matter annihilation occuring predominantly in dense clumps will produce a feature
in the local electron spectrum, that deviates from that expected if the dark matter were smoothly 
distributed. The sharp cut-off in the contribution from Kaluza-Klein dark matter is often smoothed 
out, and the spectral feature would be indistinguishable from a pulsar source,
even if the energy resolution of the electron 
detector were perfect. The spectral shape of the electron excess is insufficient to discriminate
a dark-matter origin from more conventional astrophysical explanations, contrary to a recent claim
\cite{hh}.
\item While the mass of the dark-matter particle may be misestimated by only 20\% or so, the
amplitude of the electron excess can vary by more than a factor of 2 for a clump density
$n_c = 10\ {\rm kpc^{-3}}$, and the required 
boost factors will be misestimated by the same factor.
\item All variations in the amplitude and spectral shape of the dark-matter contribution to the
local electron flux depend on the density of dark-matter clumps, the variations being larger for
smaller clump densities.
\end{itemize}
If the clump density is the decisive parameter determining the amplitude of spectral variations in
the electron excess from dark-matter annihilation, then it is of prime interest to estimate that number.
Dark-matter clumpyness provides the boost factors required by the ATIC data, which may be further increased 
by Sommerfeld corrections \cite{profumo,lattanzi}. Simulations of structure 
formation in cold-dark-matter cosmologies show clumping on a variety of scales, but the
boost factors are generally very moderate \cite{diemand,lavalle}. The clump density in those simulations is 
generally smaller than assumed in this paper, but that may be due to limited numerical resolution. The
density of simulated particles in Via Lactea II barely exceeds $10^3\ {\rm kpc^{-3}}$ at the solar circle,
and therefore the clump density will unavoidably be much lower. In any case, if the clump density is indeed
significantly lower than $n_c = 10\ {\rm kpc^{-3}}$, then the dark-matter scenario can also not be 
distinguished from a pulsar origin by studying the high-latitude
diffuse gamma-ray emission from the excess electrons, because the gamma-ray intensity distribution in 
the dark-matter scenario becomes similarly patchy as in the pulsar case.

\bibliography{atic-mkp}

\end{document}